\documentclass[preprint,review,12pt]{elsarticle}

\usepackage{amssymb}

\journal{Physica A: Statistical Mechanics and its Applications}

\begin{document}

\begin{frontmatter}


\title{Critical dynamics in systems controlled by fractional kinetic equations}



\author[bat]{L. A. Batalov}
\ead{zlokor88@gmail.com}
\author[zubr]{A. A. Batalova}
\ead{AsjaStall@gmail.com}
\address[bat]{Department of Statistical Physics, State University, Ulyanovskaya, 1, 198504, St.Petersburg , Petrodvorets, 
Russia}

\address[zubr]{Department of Atmospheric Physics, Saint-Petersburg State University, Ulyanovskaya, 1, 198504, St.Petersburg , Petrodvorets, 
Russia}

\begin{abstract}
The article is devoted to the dynamics of systems with an anomalous scaling near a critical point. The fractional stochastic equation of a Lanvevin type with the $\varphi^3$ nonlinearity is considered. By analogy with the model A the field theoretic model is built, and its propagators are calculated. The nonlocality of the new action functional in the coordinate representation is caused by the involving of the fractional spatial derivative. It is proved that the new model is multiplicatively renormalizable, the Gell-Man-Low function in the one-loop approximation is evaluted. The existence of the scaling behavior in the framework of the $\varepsilon$-expansion for a superdiffusion is established.
\end{abstract}


\end{frontmatter}


\section{Introduction}
The models based on the stochastic  processes with continuous time are widely used in modern 
theories of phase transitions, an atmospheric turbulence and electromagnetism \cite{1},\cite{2},\cite{eps_frac_der}. From the thermodynamic point of view these models are characterized by the existence of the temperature $T_{c}$ such that observed experimental results for $T \rightarrow T_c-0$ and $T \rightarrow T_c+0$ differ significantly \cite{termo}.

The object of our investigation are systems governed by stochastic equations in the vicinity of the temperature $T_{c}$ called the critical point. Problems of an equilibrium statistical physics and a thermodynamics without an explicit time-dependence are called static. Obviously in such systems there are processes the description of energy-dissipation. The description of a critical behaviour in the dynamic models is based on the analogous static case, but it is more complex technically \cite{A_model}. In critical dynamics  only the models of a Langevin type are considered, i.e. the models with the first time derivative on a field.

Among these models the Ginzburg-Landau dynamic model (the model A in the Galperin and Hoenberg classification scheme \cite{sheme}) proposed by Landau and Khalatnikov for description of a superfluidity is the most interesting . The critical dynamics of the model A was investigated  using a dimensional regularization and RG-methods based on the $\varepsilon$-expansion \cite{main_book_c}. The RG method allows to prove the existence of the critical scaling (it is a power-law dependence near a critical point of dynamic correlators) and to obtain corrections for the main exponent \cite{Vasiliev}. 

For the isotropic system which consists of particles moving under the influence of the random Gaussian distributed forces the base of theoretical analysis is the diffusion equation.

For a classical diffusion in the $d$ - dimensional space a particle travels for the time period $t$ the distance $\lambda t^{d\cdot 1/2}$, where $\lambda$ is the Onsager kinetic coefficient, which here can be understood as the diffusion coefficient. This  dependence is called a normal scaling. However, the power-law dependence of the displacement with the critical exponent more or less than  1/2 (an anomalous scaling) is observed in some physical systems \cite{frac_L}.

For such FDE the corresponding scaling exponent is equal $\alpha / \sigma$, and fractional derivatives are the generalization of integer derivatives. It allows to apply for a solution  of linear equations standard methods the Laplace and Fourier transforms. 

In the particular case $\alpha = 1,~\sigma = 2$ we obtain the standard equations of the Brownian motion. An exact solution  of the linear fractional diffusion equation was found for many processes. Some of them generalise the solutions of the standard  diffusion equation with the  exponential functions replaced by the Mittag-Leffler function \cite{diffusion}. Moreover the class of exact solutions of non-linear diffusion equations and the Fokker-Plank equations expanded very much last time \cite{Ant_new, fokker}.

 The RG methods and the ideas about description anomalous scaling with fractional differential equations (FDE) with derivative orders $\alpha$ (a time derivative) and $\sigma$ (a spatial derivative)  were developed in the same time. 
 
In our opinion the application of the renormalization group to the solution fractional differential equations will produce good results.This suggestion bases on the observation, which shows us, that much of the systems demonstrated the critical behaviour have the scaling functions with anomalous exponents. Governing equations of these systems are fractional (for example a developed turbulence \cite{ren_frac}). In our paper the technique of the RG method generalisation on dynamic models with fractional spatial and time derivatives is proposed. 

We have developed a technique of a renormalization for an arbitrary stochastic problem with the $\varphi ^m$ nonlinearity. The one-loop diagram for the most essential model of the critical dynamics - the model was calculated. We showed, that an analogy with statics took place only for $\alpha =1$ and the independence on $\sigma$ and $m$.

This technique  is convenient in applications, it is used in the calculation of higher orders, because it is based on the well-developed quantum-field renormalization-group theory \cite{Ant_vas}. However,  it will be showed in the paper, that for the general analysis and determination  of critical regimes the 1-loop approximation is enough. Before now the RG method was applied to a nonequilibrium system usually with an anomalous diffusion in the Wilson  recursive equations form \cite{A_model},\cite{france}. We emphasize, that  in our work it was used the quantum-field renormalization. The critical scaling will be proved, the main exponential function and their dependence on $\alpha$ and $\sigma$ will be found. 

The article construct as follows: in section 2  the stochastic task of the critical dynamics is manifested, and its formulation is given in the field theoretic terms. The section 3 is devoted to the calculation of propagators of this model. In section 4 the model is analysed thoroughly and a transition to the renormalized functional of action in the logarithmic dimension of space is carried out.
Below we derived the RG-equations and calculated the Gell-Mann-Low function in the one-loop approximation.





\section{Formulation of problem}

We consider a nonequilibrium system described of the scalar function  $\varphi({\bf{x}},t)$, where $t$ is time, ${\bf{x}}$ - $d$ - dimensional radius vector. All calculations demonstrated in this section are easy generalized on the case, where field $\varphi$ is multi-components.

The critical dynamic of field $\varphi({\bf{x}},t)$ is described by the stochastic Langevin equation:
\begin{equation}
\partial_t \varphi = \lambda V(\varphi)+F \label{Lang}
\end{equation}
where $V(\varphi)=\delta \bar{S} / \delta \varphi$ - variational derivative of the static action functional $\bar{S}(\varphi)$ of time-independence free field $\varphi({\bf{x}})$, $\lambda > 0$ - Onsager transport coefficient, $F$ - random external force. For this a Gaussian distribution with zero - mean  $\langle F({\bf{x}}) \rangle = 0$ and covariance
\begin{equation}
\langle F({\bf{x}},t)F({\bf{x'}},t')\rangle = 2\lambda \delta({\bf{x}}-{\bf{x'}})\delta(t-t') 
\label{power}
\end{equation}
is assumed.

The equality of the double Onsager coefficient and the correlator power provides an agreement with the static case described by the functional $\bar{S} (\varphi)$. It is consequence of the Fokker-Planck equation for the field distribution function $P(\varphi)$ \cite{main_book_c}.

The functional potential $V(\varphi)$ includes a linear differential operator of the second order (Laplacian) for different generalizations of the Ginzburg-Landau functional (further - the $\varphi ^4$ model). Therefore the linear part of the stochastic equation (\ref{Lang}) is a linear combination of the first derivative time operator and the second derivative space operator \cite{sheme}. 

The normal scaling is concluded from Eq.~(\ref{Lang}) in the leading order $g_0 = 0$ and $V(\varphi) = \Delta \varphi$: $\langle {\bf x} ^2  \rangle = \lambda t^{2dH}$, where $H$, called the Hurst exponent, which is equal $1/2$.

We can use Eq. ~(\ref{Lang}) included instead of the partial time and spatial derivatives the fractional derivatives of orders $\alpha$ and $\sigma$ related to the Hurst exponent $H=\alpha / \sigma$  for description of the anomalous diffusion processes with same parameter $\lambda$.  For representation of subdiffusion processes $0<H<1/2$ the partial time derivative could be changed by the fractional derivative of order $\alpha$ ($0 < \alpha <1$), and in the case of superdiffusion $1/2<H<1$ laplacian could be changed by the differential operator of order $\sigma$. In models of turbulent transport the Hurst exponent is greater than $1/2$ and is close to $3/4$.

Therefore we will find so generalization of the problem (\ref{Lang}), that allows to use the fractional derivatives. It is convenient to use the time derivative of order $\alpha$ $ _a^CD_t^\alpha $ in the Caputo representation with the lower terminal $a$, which is usually chosen from the condition of zero antiderivative \cite{main_book_f}:
\begin{equation}
 {_a^CD_t^\alpha} f(t)=\frac{1}{\Gamma(n-\alpha)}\int_a^t \frac{f^{(n)}(\tau)d\tau}{(t-\tau)^{\alpha+1-n}} \label{caputo}
\end{equation}
where $n$ is the nearest integer bigger than $\alpha$. This representation provides the Laplace transform as following:
\begin{equation}
\mathcal{L} \{_a^CD_t^\alpha f(t)\} = s^\alpha \mathcal{L}\{f\}+\sum\limits_{m=0}^{n-1} s^{\alpha -1 -m} \frac{\partial^m f}{\partial t^m} |_{t=0} \label{laplace}
\end{equation}
As a generalization of the Laplacian, it is convenient to define the operator in momentum representation, in other words, we will work with its Fourier image. The word "momentum" here can be understood as a wave vector.

If $\sigma$ - order of the spatial derivative and $(-\Delta)^{\sigma /2}$ - the so-called fractional Laplacian acting on the function of the Schwartz class \cite{laplace} $f$, so that $(-\Delta)^{\sigma /2} f = h$,  then
\begin{equation}
\hat{h}({\bf{k}})=k^\sigma \hat{f}({\bf{k}}),~~ k\equiv |{\bf{k}}|\label{laplace_fourier}
\end{equation}

If $0 < \sigma < 2$ we can also use the integral representation \cite{laplace}:

\begin{equation}
(-\Delta)^{\sigma /2} f({\bf{x}}) = C(n,\sigma) P.V. \int\limits_{\mathbf{R^d}} \frac{f({\bf{x}}) - f({\bf{y}})}{ \Vert {\bf{x}}-{\bf{y}} \Vert ^{d+\sigma} } d{\bf{y}} \label{laplace_int}
\end{equation}
where P.V. stands for principal value and $C(n,\sigma)=2^{\sigma -1} \sigma \Gamma ((d+\sigma)/2)/(\pi^{d/2}\Gamma (1-\sigma /2))$ is a normalization constant. Space $\mathbf{R^d}$ has standard euclidean norm $||...||$.

This allows to apply to diffusion equation Laplace or Fourier transforms. The diffusion equation arising in this way is exactly solvable model \cite{diff_new}.
\begin{equation}
_0^cD^\alpha_t \varphi({\bf{x}},t) = - \lambda (-\Delta)^{\sigma /2} \varphi({\bf{x}},t) \label{frac}
\end{equation}
supplemented by the following initial and boundary conditions:
\begin{equation}
\varphi({\bf{x}},0)=\delta({\bf{x}}),~\lim_{|{\bf{x}}| \rightarrow \infty}\varphi({\bf{x}},t)=0,~t>0  \label{correl}
\end{equation}
where $\lambda > 0$ - the diffusion coefficient, $0 < \alpha <1$,~$0 < \sigma <2$,~$\varphi({\bf{x}},t)$ - scalar field, $_0^cD^\alpha_t$ - the defined on positive half-line Caputo derivative of order $\alpha$.

The Green function of Eq. (\ref{frac}) with the initial and boundary conditions (\ref{correl}) in the time - momentum representation  is equal \cite{diffusion},\cite{diff_new}
$E_\alpha(-\lambda k^\sigma t^\alpha)$
where $E_\alpha(z)$ - the Mittag-Leffler function, which is defined usually by the following series:
\begin{equation}
E_\alpha (z)=\sum_{n=0}^\infty\frac{z^n}{\Gamma(\alpha n+1)}
\end{equation}
In the case of real and positive $\alpha$ the series converge for all values of the argument $z$, thus the Mittag-Leffler function is an entire function.

There is the simple and universal technique to transform the stochastic equation ~(\ref{Lang}), (\ref{power}), (\ref{correl}) to a field theory by doubling the number of fields \cite{double_field}. It yields an explicit form of action functional $S(\Phi)$, where 
$\Phi = (\varphi , \varphi')$ is a new two-component field. We plan to generalise this technique on the fractional derivative problem. Note, that the translational invariant perturbation theory with zero condition for $\varphi$, if $t \rightarrow -\infty$ is considered in critical dynamics, than time integration is performed on the entire axis in all terms. The initial condition (\ref{correl}) means, that integration is performed on the values $t \geq 0$ only. 

We will consider the following generalization of Eq.~(\ref{Lang})
\begin{equation}
_0^cD^\alpha_t \varphi = \lambda V(\varphi)+F \label{1}
\end{equation}
with a non-linear potential of the nonlinearity degree $m>0$:
\begin{equation}
V(\varphi)=-(-\Delta)^{\sigma /2} \varphi -\tau_0 \varphi - g_0 \varphi ^m /m! \label{2}
\end{equation}
with the initial and boundary conditions
\begin{equation}
\lim_{|{\bf{x}}| \rightarrow \infty}\varphi({\bf{x}},t)=0,~\varphi({\bf{x}},-\infty) = 0 \label{3}
\end{equation}
and the conditions on orders of derivatives:
\begin{equation}
0<\alpha<1,~~
0<\sigma<2 \label{first_res}
\end{equation}
and we assume the positivity of temperature and coupling constants: $\tau_0 >0,~g_0>0$.

The power of correlation function (\ref{power}) remains the same and it is equal $2\lambda$, because in the proof of static and dynamic theory consistency the condition $ V (\varphi) = \delta \bar{S} / \delta \varphi $ is used \cite{Ant_new}. It will be proved later. In the case of the Cauchy problem (\ref{correl}) the equilibrium distribution is realized only in the asymptotic limit $ t \rightarrow + \infty $, in which there is no dependence on time and on the initial data.

The potential $V(\varphi)$ from (\ref{1}) can be written as $V(\varphi) = \delta   \bar{S}(\varphi) / \delta \varphi$, where the static action functional is 
\begin{equation}
 \bar{S}(\varphi)  = -\frac{1}{4} P.V. \int\limits_{\mathbf{R^d} \times \mathbf{R^d}} \frac{(\varphi({\bf{x}})-\varphi({\bf{y}}))^2}{\Vert{\bf{x}}-{\bf{y}}   \Vert ^{d+\sigma}}d{\bf{x}}d{\bf{y}} -\int\limits_{\mathbf{R^d}} \left( \frac{\tau_0}{2}\varphi({\bf{x}})^2 + \frac{g_0}{(m+1)!}\varphi({\bf{x}})^{m+1} \right) d{\bf{x}} \label{stat}
\end{equation}
in the $d$ - dimensional coordinate space with coupling constant $g_0$. For proof of the statement $V(\varphi) = \delta   \bar{S}(\varphi) / \delta \varphi$ it is enough to write the difference $\bar{S}(\varphi+\delta \varphi)-\bar{S}(\varphi)$ and to take into account of the symmetry with respect to values ${\bf{x}}$ and ${\bf{y}}$ in (\ref{stat}). It's important, that the first term in (\ref{stat}) is nonlocal, i.e. which is not polynom or function of the field or its derivative evaluated at a single point in the space. This reflects physically the existence of a long-range interaction in the system, and confirms the choice of operator generalization of the spatial derivative (\ref{laplace_int}). Near a critical point all interactions are long range, it is confirmed by an experimental data. 

Application of the renormalization group method proposes the transforming of some parameters on renormalized. The non-renormalized parameters are marked by the subscript "0" to distinguish it from corresponding renormalized. The functional (\ref{stat}) at $m=3$ is nonlocal generalization of the Ginzburg-Landau functional, in the field theory terms it is the massive $\varphi ^4$ model with mass $\sqrt{\tau_0}$.

For the renormalization of Eq. (\ref{1}) we will use the theorem of equivalence of stochastic dynamic and quantum field theory with two-component field with the main field $\varphi({\bf x},t)$ and the auxiliary field $\varphi'({\bf x},t)$ \cite{double_field}. In the proof of this theorem assumption about just first-order of time derivative was not used. It was required only linearity of time-dependent operator. Therefore Eq. (\ref{1}) is equivalent to field - theoretic model with action
\begin{equation}
S(\Phi) =\int\limits_{\mathbf{R^d}} d^d{\bf{x}} \int\limits_{-\infty}^{+\infty} dt \left( \lambda_0 \varphi' \varphi' + \varphi' \left(-{^c_0D^\alpha_t} \varphi +\lambda_0 V(\varphi)   \right)\right) \label{action}
\end{equation}
where $\Phi=(\varphi,\varphi')$. For the Cauchy problem (\ref{correl}) the action functional is \cite{Ant_new}:
\begin{equation}
S(\Phi) =\int\limits_{\mathbf{R^d}} d^d{\bf{x}} \int\limits_{0}^{+\infty} dt \left( \lambda_0 \varphi' \varphi' + \varphi' \left(-{^c_0D^\alpha_t} \varphi +\lambda_0 V(\varphi)  + \delta ({\bf{x}})\delta (t-t') \right)\right) \label{correl_action}
\end{equation}
that is reduced to adding to the action of term $\varphi'({\bf{x}}=0,t=0)$ ,since always $t>t'$. Statistical correlation functions (Green functions) are defined by functional average with weigh $\exp S(\Phi)$ (we include in $S$ the usual minus sign in the argument ). Further we will work directly with the field - theoretical model (\ref{action}). We are concerned with the restriction of parameters $\alpha, \sigma, d$, allowing to apply the renormalization group method. 

The asymptotics of correlators and other quantities at $\tau \rightarrow +0$ with already the renormalized parameter $\tau$ is called the critical asymptotics \cite{main_book_c}. The physical meaning of this parameter is the deviation of the critical temperature: $\tau \sim T-T_c$. Our model at $\alpha \rightarrow 1$, $\sigma \rightarrow 2$, $m \rightarrow 3$ transforms to the model A, where unique non-trivial critical asymptotic is dependence of relaxation time on $\tau$. It is determined by the critical exponent $z$ and has the form $\tau ^{2+z}$. Other critical asymptotics coincide with those in the static $ \varphi ^ 4 $.model.  For arbitrary $\alpha$, $\sigma$ and $m$ it is false. In following section the difference between statics and dynamics for the field - theoretical model (\ref{action}) with fractional derivatives for arbitrary vertex degree $m$ will be shown.

\section{Determination of propagators}
Standard Feynman diagram method with vertex $\varphi ' \varphi ^3$ and bare propagators is corresponded to action (\ref{action}).
These propagators are found by a representation of the quadratic (called "free") term of the action 
\begin{equation}
S_0(\Phi) =\int\limits_{\mathbf{R^d}} d^d{\bf{x}} \int\limits_{-\infty}^{+\infty}dt \left(\lambda_0 \varphi' \varphi' + \varphi' \left(-{^c_0D^\alpha_t} \varphi -\lambda_0(-\Delta)^{\sigma/2} \varphi -\lambda_0\tau_0 \varphi   \right)\right) \label{free}
\end{equation}
 as the integral of a quadratic form with a symmetric operator matrix $K\in M(2,2)$:
\begin{equation}
 S_0(\Phi)=-\frac{1}{2}\int\limits_{\mathbf{R^d}} d^d{\bf{x}} \int\limits_{-\infty}^{+\infty} dt \Phi K \Phi
 \end{equation}
 where $\Phi = (\varphi , \varphi ')$. Then $ K^{-1}({\bf{x}}-{\bf{x'}},t-t') $ is the translationally invariant matrix of functions such that $ KK ^ {-1} = I \delta ({\bf{x}}-{\bf{x'}}) \delta (t-t') $, where $I$ is the identity matrix. Using integration by parts rule for fractional derivative with respect to $t$, we obtain from (\ref{free}) the following matrix expression of matrix  $ K $ in the momentum representation (let $\Xi = \lambda_0(k^{\sigma}+\tau_0)$):

\begin{equation}
\left( \begin{array}{ccc}
0 & _t^cD^{\alpha}_0+\Xi  \\
 _0^cD_t^{\alpha}+\Xi  &2\lambda_0  \\
 \end{array} \right) \label{operator}
\end{equation}
where $_t^cD^{\alpha}_0$ is right Caputo derivative defined as
\begin{equation}
 {_t^cD^{\alpha}_a} f(t)=\frac{(-1)^n}{\Gamma(n-\alpha)}\int_t^a \frac{f^{(n)}(\tau)d\tau}{(\tau-t)^{\alpha+1-n}} \label{caputo_right}
\end{equation}
We represent $ K ^ {-1} $ as
\begin{equation}
\left( \begin{array}{ccc}
\mathcal{G}_{11}({\bf{x}}-{\bf{x'}},t-t') & \mathcal{G}_{12}({\bf{x}}-{\bf{x'}},t-t')  \\
\mathcal{G}_{21}({\bf{x}}-{\bf{x'}},t-t') & 0  \\
\end{array} \right) \label{propagator}
\end{equation}
where the functions $\mathcal{G}_{11}$ and $\mathcal{G}_{12}$ called propagators are sought in the class of functions vanishing at $\Vert {\bf{x}}-{\bf{x'}}  \Vert \rightarrow \infty$. For correctness of the iteration procedure for the equation (\ref{1}) and respectively the diagram technique it is also necessary, that the condition of stability $\mathcal{G}_{12} \rightarrow 0$ at $t-t' \rightarrow +\infty$ is complied. Typically, the propagator is normalized to delta - function at coinciding times, and the requirement of delay corresponding to the physical principle of causality is imposed \cite{main_book_c, double_field}. It can be expressed by
\begin{equation}
\mathcal{G}_{12}({\bf{x}}-{\bf{x'}},t-t') = \left\{ \begin{array}{rl}
\delta({\bf{x}}-{\bf{x'}}) &\mbox{ if $t=t'+0$} \\
0 &\mbox{ if $t<t'$} \label{initial}
\end{array} \right.
\end{equation}
From this we see that to find the propagator $\mathcal{G}_{12}$ we can use the Laplace transform in the variable $t-t'$ to the variable $s$ and the Fourier transform in the variable ${\bf x-\bf x'}$ to the variable ${\bf k}$. For convenience, different representations of propagators  differ in this text by their arguments.
Then the equation $ KK ^ {-1} = I \delta ({\bf{x}}-{\bf{x'}}) \delta (t-t') $ with (\ref{operator}) and (\ref{propagator}) - (\ref{initial}) can be rewritten as follows at $t>0$:
\begin{equation}
s^\alpha \mathcal{G}_{12}({\bf{k}},s)-s^{\alpha - 1}\mathcal{G}_{12}({\bf{k}},+0)+\Xi \mathcal{G}_{12}({\bf{k}},s) \label{eq_prop_12}
\end{equation}

We use the property of the Mittag-Leffler function $E_\alpha (ut^\alpha)$ for $Re(s)>|u|^{1/\alpha}$:

\begin{equation}
\int_0^{\infty} e^{-st}  E_\alpha (ut^\alpha)dt = \frac{s^{\alpha -1}}{s^\alpha - u} \label{laplace_mittag}
\end{equation}

We finally obtain from (\ref{initial}) - (\ref{laplace_mittag}):
\begin{equation}
\mathcal{G}_{12}(t-t',k)=\theta(t-t')E_\alpha \left( 
-\Xi (t-t')^\alpha
 \right) \label{G_12}
\end{equation}

Similarly, we obtain the advanced propagator:

\begin{equation}
\mathcal{G}_{21}(t-t',{\bf{k}})=\theta(t'-t)E_\alpha \left( 
-\Xi (t'-t)^\alpha
 \right) \label{G_21}
\end{equation}
Note, that the propagator (\ref{G_12}) coincides with solution of the Cauchy problem (\ref{frac}), supplemented by the condition $t>0$.
Just as in model A, $\mathcal{G}_{12}$ is the retarded propagator and $\mathcal{G}_{21}$ is advanced. The propagator $\mathcal{G}_{11}$ can be uniquely determined by the convolution of functions $\mathcal{G}_{12}$ and $\mathcal{G}_{21}$. From (\ref{operator}),(\ref{propagator}) we obtain:
\newpage
\begin{eqnarray}
\mathcal{G}_{11}({\bf{k}},t) = 2\lambda_0 \int\limits_{-\infty}^{+\infty} \mathcal{G}_{12}({\bf{k}},t-t')\mathcal{G}_{21}({\bf{k}},t')dt =\\
= 2\lambda_0 \int\limits_{-\infty}^{+\infty} \theta (t-t')\theta(-t') E_\alpha (-\Xi (t-t')^\alpha)E_\alpha(-\Xi (-t')^\alpha)dt =\\
=2\lambda_0 \left(\theta(-t)\int\limits_{-\infty}^{t}E_\alpha (-\Xi (t-t')^\alpha)E_\alpha(-\Xi (-t')^\alpha)dt+\theta(t)\int\limits_{-\infty}^{0}E_\alpha (-\Xi (t-t')^\alpha)E_\alpha(-\Xi (-t')^\alpha)dt \right) = \\
=2\lambda_0\int\limits_{0}^{+\infty}E_\alpha (-\Xi t'^\alpha)E_\alpha(-\Xi (t'+|t|)^\alpha)dt=\\
=2\lambda_0\Xi^{-1/\alpha}\int\limits_{0}^{+\infty}E_\alpha (-t'^\alpha)E_\alpha(- (t'+|t|\Xi^{1/\alpha})^\alpha)dt
 \label{convolution}
\end{eqnarray}

Usually the propagators $\mathcal{G}_{11}$ and $\mathcal{G}_{12}$ are derived in the frequency - momentum representation. If $\alpha$ is fractional, the time derivative is replaced by factor $(i\omega)^\alpha$, this factor fixes a single-valued branch. This is a well-known problem of non-physical poles separation. In our view it is more convenient to use the $({\bf k},t)$ - representation.

\section{Renormalization of field theoretical model in one-loop approximation}
\subsection{The analysis of scale invariance}
We consider in this section the more general class of models with power-law nonlinearity $\varphi^m$:
\begin{equation}
S(\Phi) = S_0(\Phi)-\int\limits_{\mathbf{R^d}} d^d{\bf{x}} \int\limits_{-\infty}^{+\infty} dt \frac{g_0\lambda_0\varphi^m}{m!} \label{m-act}
\end{equation}

It is convenient to do stretching variables in the auxiliary field $\varphi' \rightarrow \varphi' c_0$ with $c_0=\lambda_0^{-1}$. Than the action (\ref{m-act}) takes the form:
\begin{equation}
S(\Phi) = \int\limits_{\mathbf{R^d}} d^d{\bf{x}} \int\limits_{-\infty}^{+\infty} 
\left(
c_0 \varphi' \varphi' +\varphi'
\left(
-c_0{^c_0D^\alpha_t} \varphi -(-\Delta)^{\sigma/2} \varphi -\tau_0 \varphi -\frac{g_0\varphi^m\varphi'}{m!}
\right)
\right) \label{str-action}
\end{equation}

We will consider the theory in the coordinate space of dimension $d$ .Further it is necessary to analyse scaling dimensions. We introduce the definition of two independent canonical dimensions: the momentum dimension $\mu^p$, the frequency dimension $\mu^\omega$ and total for an arbitrary value $a$  $\mu_a=\alpha \mu^p_a +\sigma \mu^\omega_a$ \cite{Ant_vas}. By definition, $\mu_p^p=-\mu_x^p=\mu^\omega_\omega=-\mu_t^\omega =1$,~$\mu_\omega ^p=\mu_t^p=\mu_p^\omega=\mu_x^\omega=0$, and dimensions of other quantities in (\ref{action}) are found requiring dimensionless of each term in the action. From this follows, that scaling dimension of integration over all space and time is equal $\sigma + \alpha d$, scaling dimension of time derivative is $\sigma$, the scaling dimension of spatial derivative is $\alpha$. Then the system of equations, expressed dimensionless of the action (\ref{action}) is
\begin{eqnarray}
\mu_{c_0}+2\mu_{\varphi'}=\alpha d +\sigma \\\nonumber
\mu_{c_0}+\mu_\varphi+\mu_{\varphi'}+\alpha \sigma =\alpha d +\sigma\\\nonumber
\mu_\varphi+\mu_{\varphi'}+\alpha \sigma =\alpha d +\sigma\\\nonumber
\mu_\varphi+\mu_{\tau_0}+\mu_{\varphi'} = \alpha d +\sigma\\
\mu_{\varphi'}+\mu_{g_0}+m\mu_\varphi= \alpha d +\sigma
\label{dim-less}
\end{eqnarray}
Considering jointly the second and third equations in (\ref{dim-less}) we get $\mu_{c_0} =0$, from third and fourth we obtain $\mu_{\tau_0} =\alpha \sigma$. We obtain immediately from first equation in (\ref{dim-less}) dimension of the auxiliary field $\mu_{\varphi'} = (\alpha d +\sigma)/2$. Dimension of the main field is obviously $\mu_{\varphi} = (\alpha d +\sigma)/2-\alpha \sigma$. Finally, the fifth equation follows $2\mu_g=\alpha d (1-m)+\sigma (1-m) +2\alpha \sigma m$.

 Further, the theory is considered in the space, where the coupling constant  $g_0$ becomes dimensionless. In this case existing divergences in correlators, if  the condition of multiplicative renormalizability is complied, could be manifest in the form of poles at the deviation of the space dimension $\varepsilon$. This space dimension  $d^*$ is called logarithmic. In our case
\begin{equation}
d^* = \sigma \left(\frac{2m}{m-1}-\frac{1}{\alpha}\right) \label{log}
\end{equation}
Despite the possibility of analytic continuation of diagrams in the region $d<0$, we think, that it is not correspond to the physical meaning of the problem. Equation ~(\ref{log}) clarifies restrictions on parameter $\alpha$ as condition on positiveness of spatial dimension.   Considering ~(\ref{log}), they become
\begin{equation}
\frac{m-1}{2m}<\alpha<1 \label{restrict}
\end{equation}
This inequality (\ref{restrict}) along with the condition of the fractional representation of the Laplacian as the principal value of the integral operator (\ref{laplace}) and (\ref{first_res}) $1<\sigma <2$ is conditions for the applicability of the quantum-field renormalization group method to the equation (\ref{1}). Hence we see as increasing degree of nonlinearity $m$ in the action narrows the range of permissible $\alpha$. The equation (\ref{log}) expresses the relation between the dimension of the coordinate space $d$ and the orders of derivatives $\alpha$ and $\beta$. This provides a self-consistency of theoretical-field model (\ref{m-act}). From inequality $(m-1)/2m < 1/2$ implies that for the usually considered range $1/2 < \alpha <1$ always satisfies (\ref{restrict}).

Now for applying the renormalization group method in the space dimension  $d=d^*-\varepsilon$ it is necessary to prove the multiplicative renormalizability of the theory \cite{Ant_vas}. The form of required counterterms is found using scaling analysis of 1-irreducible Green functions. Such functions with $N_\Phi \equiv N_\varphi , N_{\varphi '}$ external legs have scaling dimension $\delta = \alpha d+\sigma-\mu_\Phi N_\Phi = \alpha d+\sigma -\mu_{\varphi'}N_{\varphi'}-\mu_{\varphi}N_{\varphi}$, where $\mu_\Phi$ is field scaling dimension. The divergence is determined by the total canonical dimension $\delta$: the diagram has surface divergence, if in logarithmic theory $\delta$ is integer non-negative number. In our case at $\varepsilon = 0$ we find from (\ref{dim-less}), (\ref{log})
\begin{equation}
\frac{\delta (m-1)}{\sigma \alpha} = 2m-N_{\varphi}-mN_{\varphi'} \label{div_ind}
\end{equation}
This means that surface divergences exist only in 1-irreducible functions $ \langle \varphi' \varphi' \rangle $, $ \langle \varphi '\varphi \rangle $ and $ \langle \varphi' \varphi ^ m \rangle $, i.e. only in the functions presented in (\ref{action}).

Now we can write the renormalized action functional
\begin{equation}
S(\Phi) = \int\limits_{\mathbf{R^d}} d^d{\bf{x}} \int\limits_{-\infty}^{+\infty}
 \left( Z_1 c \varphi' \varphi' + \varphi' \left(-Z_2c~ {_0^cD^\alpha_t} \varphi -Z_3 (-\Delta)^{\sigma/2} \varphi -Z_4 \tau \varphi -Z_5\frac{g M^\varepsilon \varphi^m}{m!}   \right)\right) \label{ren_action}
\end{equation}
with renormalization mass $M$, renormalized parameters $c,~\tau ,~g$ and relations between dimensionless (they depends on $g$ only) renormalization constants $Z$:
\begin{eqnarray}
\nonumber
Z_1=Z_{\varphi'} Z_c\\ \nonumber
Z_2=Z_{\varphi'}^{1/2}Z_\varphi ^{1/2} Z_c \\ \nonumber
Z_3=Z_{\varphi'}^{1/2}Z_\varphi ^{1/2}\\ \nonumber
Z_4=Z_{\varphi'}^{1/2}Z_\varphi ^{1/2}Z_\tau \\ 
Z_5= Z_{\varphi'}^{1/2}Z_\varphi ^{m/2}Z_g  \label{constants}
\end{eqnarray}

It is known, that in model A there is a single new renormalization constant $Z_c$ (compared with the static case). Equality of other constants follows from uniformity of Schwinger equation for static and dynamic generating functionals. The proof of this is based on $V(\varphi) = \delta \bar{S}(\varphi)/\delta \varphi$, where $\bar{S}(\varphi)$ is the renormalized static functional.

\begin{equation}
 \bar{S}(\varphi)  = -\bar{Z}_\varphi \frac{1}{4} P.V. \int\limits_{\mathbf{R^d} \times \mathbf{R^d}} \frac{(\varphi({\bf{x}})-\varphi({\bf{y}}))^2}{\Vert{\bf{x}}-{\bf{y}}   \Vert ^{d+\sigma}}d{\bf{x}}d{\bf{y}} -\int\limits_{\mathbf{R^d}} \left( \bar{Z}_\tau \bar{Z}_\varphi \frac{\tau}{2}\varphi^2 + \bar{Z}_g\bar{Z}_\varphi^{(m+1)/2}\frac{gM^\varepsilon}{(m+1)!}\varphi^{m+1} \right) d{\bf{x}} \label{stat_ren}
\end{equation}
In general case we have following 
\newtheorem{thm}{Theorem}
\begin{thm}
The equality of the renormalization constants (\ref{stat_ren}) and (\ref{ren_action}) 
\begin{equation}
Z_\varphi = Z_{\varphi'}=\bar{Z}_\varphi,~Z_g=\bar{Z}_g,~Z_\tau=\bar{Z}_\tau
\end{equation}
holds if and only if $\alpha = 1$.
\end{thm}

The proof is constructed like model A, taking into account the fact that the logarithmic dimension of the static action functional (\ref{stat}) is
\begin{equation}
d^*_{stat} = \sigma \frac{m+1}{m-1} \label{log_stat}
\end{equation}
Obviously, the coincidence of the logarithmic dimension (\ref{log_stat}) and (\ref{log}) takes place only at $\alpha = 1$.
Thus, all UV (at large momentum) divergences manifest themselves as poles at $\varepsilon$ in the renormalization constants $Z_i,~i=1...5$. Further, standard renormalization group equations expressed the renormalization invariance are written. We introduce the solution of Eq. (\ref{1}) as follows at $e_0 = (\tau_0 , \lambda_0 )$:
\begin{equation}
G({\bf{x}},t|e_0) = \int \mathcal{D}\varphi' \int \mathcal{D}\varphi \varphi({\bf{x}},t) \exp(S(\varphi' ,\varphi)) \label{Green}
\end{equation}
Here the normalization constant is included into the differential $\mathcal{D}\varphi'  \mathcal{D}\varphi$.

The condition of multiplicative renormalizability implies the relation $G(e_0 ) = Z^{-1} G_R (e, M)$ for the corresponding Green functions in Eq. (\ref{Green}). We use $\tilde{\mathcal{D}}_M$ to denote the differential operation $M\partial_M$ for fixed $e_0$ and operate on both sides of this equation with it. This gives the basic RG equation $\tilde{\mathcal{D}}_M G_R=0$:

\begin{equation}
[\mathcal{D}_M+\beta(g)\partial_g - \gamma_g]G_R(e,M)=0,~\mathcal{D}_x \equiv x\partial_x \label{rg-eq}
\end{equation}
where the anomalous dimension $\gamma_g$ in using the minimal subtraction scheme (MS) is expressed in terms of the renormalization constant $Z_g$: $\gamma _g(g) = \tilde{\mathcal{D}}_M \ln Z_g =-\varepsilon g\partial_g\ln Z_g /(1+g\partial_g\ln Z_g)$ and doesn't include poles at $\varepsilon$. We need to further analyse the expression for the Gell-Mann-Low function $\beta(g)= \tilde{\mathcal{D}}_M g$ in terms of the anomalous dimension of the coupling constant $g$:
\begin{equation}
\beta(g) = -g(\varepsilon +\gamma_g(g)) \label{common_beta}
\end{equation}
The anomalous dimension of an arbitrary value $a$ expressed in terms of the corresponding renormalization constant as $\gamma_a=\beta(g)\partial_g\ln Z_a$.

 Obviously, that only as $ \alpha = 1 $ and  the logarithmic space dimension of the dynamic theory coincides with the static case, we can use theorem for the coincidence of all renormalization constants excluding $ Z_c $ and accept  behaviour of the Gell-Mann-Low $ \beta (g) $ function of the static model \cite{Ant_vas}. In the general case of the dynamic problem (\ref{action}) we can not do it.

\subsection{Calculation of the one-loop diagrams}
\begin{figure}
\begin{center}
\includegraphics[scale = 0.5]{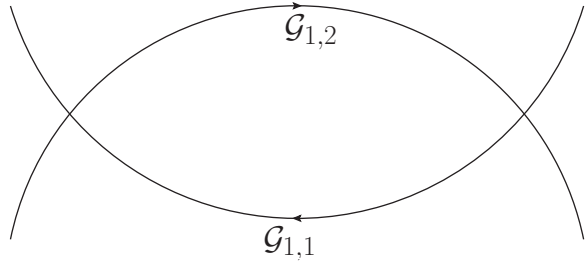}
 \caption{The only one-loop diagram, propagators $\mathcal{G}_{11}$ and $\mathcal{G}_{12}$ are signed.} 
 \label{ris:dia}
\end{center}
\end{figure}

Now we consider the case $m=3$. In one-loop approximation there is only one diagram $\varphi ' \varphi ^3$ (Fig.~\ref{ris:dia}) , which contribute the ordinary term in renormalization constants: $Z_g$,$Z_\tau$. We can calculate it in the MS-scheme at the null external momentum and the null external frequency. The condition $\omega_{ext}=0$ is equivalent to the equality of time arguments of propagators. The pole part of this diagram $\Gamma_{\varphi' \varphi ^3 }$ is denoted as  $\chi(\alpha , \sigma)$: $\Gamma_{\varphi' \varphi ^3 } \sim \chi(\alpha , \sigma) / \varepsilon$. 

In MS scheme only the function  $\chi(\alpha , \sigma)$ will interest us. For its calculation in diagram all external momentum are supposed zero. The diagram $\Gamma_{\varphi' \varphi ^3 }$ will equal:
\begin{equation} 
\Gamma_{\varphi' \varphi ^3 } = \int\limits_0^\infty dt \int\limits_0^\infty dk k^{d^*-\varepsilon-1}
\mathcal{G}_{11}(k,t)\mathcal{G}_{12}(k,t)
 \label{main_diagram} 
\end{equation}
 where the logarithmic dimension $d^*$ is defined by ~(\ref{log}). Also we omit factors independent on $\alpha$ and $\sigma$ in the diagram. The final factor is determined from a compliance with the static case.  Set in (\ref{main_diagram}) (\ref{G_12}) and (\ref{convolution}), we can submit this as an integral:
\begin{equation}
\Gamma_{\varphi' \varphi ^3 } = \int\limits_0^\infty dt \int\limits_0^\infty dk
k^{d^*-\varepsilon-1}   E_\alpha (-\Xi t^\alpha) \int\limits_0^\infty d\tau  
E_\alpha (-\tau^\alpha)E_\alpha(- (\tau+t\Xi^{1/\alpha})^\alpha)
\label{diagram}
\end{equation}

Replacing $\Xi t^\alpha \rightarrow t^\alpha$, we then obtain the product of two integrals
\begin{equation}
\Gamma_{\varphi' \varphi ^3 } = 2\lambda\int\limits_0^\infty dk
 \frac{k^{d^*-\varepsilon-1}}{\Xi ^ {2/\alpha}}   
 \int\limits_0^\infty dt \int\limits_0^\infty d\tau
 E_\alpha (-t^\alpha) E_\alpha (-\tau^\alpha) E_\alpha (-(t+\tau)^\alpha) \label{dia_now}
\end{equation}
The integral over the momentum space is now
\begin{eqnarray}
\int\limits_0^\infty dk
 \frac{k^{d^*-\varepsilon-1}}{\Xi ^ {2/\alpha}} = \tau^{\frac{1}{\sigma}(d^*-\varepsilon -1)-\frac{1}{\alpha}}\lambda ^{-\frac{2}{\alpha}}\int\limits_0^\infty dy \frac{y^{(d^*-\varepsilon)/\sigma -1}}{(y+1)^{2/\alpha}} =\nonumber\\ 
  \tau^{\frac{1}{\sigma}(d^*-\varepsilon -1)-\frac{1}{\alpha}}\lambda ^{-\frac{2}{\alpha}} 
 \frac{\Gamma (\frac{2}{\alpha}-\frac{d^*}{\sigma}
 +\frac{\varepsilon}{\sigma})\Gamma (\frac{d^*}{\sigma}-\frac{\varepsilon}{\sigma})}{\Gamma (\frac{2}{\alpha})}= \nonumber\\
 \tau^{\frac{1}{\sigma}(d^*-\varepsilon -1)-\frac{1}{\alpha}}\lambda    ^{-\frac{2}{\alpha}} \frac{\Gamma (\frac{3}{\alpha}-3+\frac{\varepsilon}{\sigma})\Gamma (3-\frac{1}{\alpha}-\frac{\varepsilon}{\sigma})}{\Gamma (\frac{2}{\alpha})} \label{pole}
\end{eqnarray}
and the diagram (\ref{diagram}) is equal
\begin{equation}
\Gamma_{\varphi' \varphi ^3 } = \tau^{\frac{1}{\sigma}(d^*-\varepsilon -1)-\frac{1}{\alpha}}\lambda    ^{-\frac{2}{\alpha}} \frac{\Gamma (\frac{3}{\alpha}-3+\frac{\varepsilon}{\sigma})\Gamma (3-\frac{1}{\alpha}-\frac{\varepsilon}{\sigma})}{\Gamma (\frac{2}{\alpha})} J(\alpha)
\end{equation}
where we denoted 
\begin{equation}
J(\alpha) = \int\limits_0^\infty dt \int\limits_0^\infty d\tau
 E_\alpha (-t^\alpha) E_\alpha (-\tau^\alpha) E_\alpha (-(t+\tau)^\alpha)
\end{equation}
In particularly at $\alpha = 1$~$E_\alpha (x) = \exp (x)$ and $J(\alpha) = 1/4$. 

The expression (\ref{pole}) has a simple pole in $\varepsilon$ at $\alpha = 1$ and at $\alpha = 1/n,~n\in \mathbb{Z},~n\geqslant 3$. However, the situation $n\geqslant 3$ is forbidden by the condition (\ref{restrict}). Therefore at $\alpha =1$ and $1<\sigma <2$   we have $\chi (1,\sigma) = \sigma$. In a range $1/3 < \alpha < 1$ the expression (\ref{pole}) is regular, thus $\chi (\alpha , \sigma) =0$. It is convenient to construct the perturbation series of the coupling constant $g' = g/S_d$, where $S_d$ - surface area of the $d$ - dimensional sphere: $S_d=2\pi^{d/2}/2^d\Gamma (d/2)$ \cite{main_book_c}. Than in the case $0<\alpha <1/3$ the renormalization constant $Z_g$ will be regular in the one-loop approximation. Hence, if 
\begin{equation}
\gamma_g(g')=\gamma_g^{(1)}g'+O(g'^2) \label{lin}
\end{equation}
then $\gamma_g^{(1)} =0$. But in the case $\alpha = 1$ we conclude 
\begin{equation}
\gamma_g^{(1)} = C\sigma \label{lin2}
\end{equation}
where $C$ is a constant which is not depend on $\sigma$. It is known, that in the model A at $d=4$ $\beta^A(g')=-g'\varepsilon +3g'^2$. Finally from (\ref{common_beta}) and (\ref{lin}),(\ref{lin2}) we conclude
\begin{equation}
\beta (g') = -g'\varepsilon +\frac{3}{2}\sigma g'^2 \label{main_res}
\end{equation}
The important consequence of the formula (\ref{main_res}) is the existence of the IR stable point of the Gell-Man-Low function $g'^* = 2\varepsilon / 3\sigma \sim \varepsilon:~\beta (g'^*)=0$. This fact confirms the critical scaling at $\alpha=1,~0<\sigma <2$ in generalized model A.

\section{Conclusion}

We applied еру field theoretic renormalization group tool to the stochastic fractional derivative nonlinear equation (\ref{1}). 
In principle, we could realize this procedure even for the nonstochastic Cauchy problem \cite{Ant_new}, but it is important for us to demonstrate a possibility of the MSR formalism application to fractional processes. Our work represents a generalization of a critical dynamics research observed in the model A. Inclusion  of fractional-order derivatives of $\alpha$ with respect to time and of $\sigma$  with respect coordinates to the model leads to some difficulties.

Replacing the standard Laplacian by the fractional analogue leads to a nonlocal both action functionals: static and dynamic. That is why the propagators are sought in the momentum representation, in which the action is local. There are not other difficulties due to the substitution $\Delta \rightarrow -(-\Delta)^{\sigma /2}$ apart a nonlocality of action, and the generalization is trivial. The case $\alpha = 1,~\sigma <2$ is called superdiffusion.

Propagators of our model at $\alpha \neq 1$ are expressed in terms of the Mittag-Leffler function $E_\alpha(x)$ depending on the complex parameter $\alpha$. The propagator (\ref{convolution}) has a complex integral representation for multi-loop calculations. This is due to the fact that the semigroup property of the Mittag - Leffler $E_\alpha (-x^\alpha)+E_\alpha (-y^\alpha) = E_\alpha (-(x+y)^\alpha)$ takes place only at $\alpha = 1$. In this case the Mittag - Leffler function coincides with the exponential function.

We prove the multiplicative renormalizability of the model with the nonlinearity of the general form $\varphi^m$ and found its logarithmic dimension (\ref{log}). In the physically interesting range $1/2 < \alpha <1$ the logarithmic dimension is always positive.

One application of the renormalization group method to the theory of phase transitions is the establishment of scaling behaviour in the infrared asymptotic region $k\rightarrow 0,~\omega \rightarrow 0$. A necessary condition for this is the fact that the Gell-Mann-Low function has a fixed point: $\beta (g^*)=0,~g^* \sim \varepsilon$. From the one-loop calculation we can see, that such situation is realised with the restrictions (\ref{restrict}) only for the superdiffusion $\alpha =1,~\sigma \leq 2$. For the subdiffusion $\alpha \leq 1,~\sigma = 2$ we can not assert something specific.

Accordingly if $\alpha = 1$ and the spatial dimension is less than $d^*=2\sigma$, then in the systems controlled by the equation (\ref{1}-\ref{3}) the critical scaling is predicted by us.

\section{Acknowledgements}
This research is supported by the Chebyshev Laboratory  (Department of Mathematics and Mechanics, St. Petersburg State University)  under RF Government grant 11.G34.31.0026

\bibliographystyle{model1a-num-names}
\bibliography{<your-bib-database>}







\end{document}